# Bipolar supercurrent in graphene

Hubert B. Heersche[1]*, Pablo Jarillo-Herrero[1]*, Jeroen B. Oostinga[1], Lieven M. K. Vandersypen[1] and Alberto F. Morpurgo[1]

[1]*Kavli Institute of Nanoscience, Delft University of Technology, PO Box 5046, 2600 GA, Delft, The Netherlands.* * *These authors contributed equally to this work.*

**Graphene – a recently discovered one-atom-thick layer of graphite[1] – constitutes a new model system in condensed matter physics, because it is the first material in which charge carriers behave as massless chiral relativistic particles. The anomalous quantization of the Hall conductance[2,3], which is now understood theoretically[4,5], is one of the experimental signatures of the peculiar transport properties of relativistic electrons in graphene. Other unusual phenomena, like the finite conductivity of order $4e^2/h$ at the charge neutrality (or Dirac) point[2], have come as a surprise and remain to be explained[5-13]. Here, we study the Josephson effect[14] in graphene[15]. Our experiments rely on mesoscopic superconducting junctions consisting of a graphene layer contacted by two closely spaced superconducting electrodes, where the charge density can be controlled by means of a gate electrode. We observe a supercurrent that, depending on the gate voltage, is carried by either electrons in the conduction band or by holes in the valence band. More importantly, we find that not only the normal state conductance of graphene is finite, but also a finite supercurrent can flow at zero charge density. Our observations shed light on the special role of time reversal symmetry in graphene and constitute the first demonstration of phase coherent electronic transport at the Dirac point.**

Owing to the Josephson effect[14, 16], a supercurrent can flow through a normal conductor (N) placed in between two closely spaced superconducting electrodes (S). For this to happen, transport in the N material must be phase coherent and time reversal symmetry (TRS) must be present. In graphene, the Josephson effect can be investigated in the "relativistic" regime[15], where the supercurrent is carried by Dirac electrons. However, it is not a-priori clear that graphene can support supercurrents, since other quantum interference phenomena that require both phase coherence and TRS were found to be absent or strongly suppressed in previous experiments[17]. Below we show experimentally that the Josephson effect in graphene is a robust phenomenon, and argue that its robustness is intimately linked to its unique electronic structure.

Single and few layer graphene Josephson junctions are fabricated on oxidized Si substrates by mechanical exfoliation of bulk graphite[1], followed by optical microscope inspection to locate the thinnest graphitic flakes, and electron beam lithography to define electrical contacts. Figure 1a shows an atomic force microscopy (AFM) image of a typical device. We use as superconducting contacts a Ti/Al bilayer (10/70nm). Titanium ensures good electrical contact to graphene and Al establishes a sufficiently high critical temperature to enable the observation of supercurrents in a dilution refrigeration set-up[18]. Before discussing their superconducting properties, we first characterize the devices with the superconducting electrodes in the normal state. Figure 1c shows the two-terminal resistance, $R$, versus gate voltage, $V_G$, for one of our samples. The strong $V_G$-dependence of $R$ provides a first indication that the device consists of at most a few layers of graphene[1], since, due to screening, $V_G$ affects the carrier density only in the bottom one or two layers. For single layers, the position of the resistance maximum corresponds to the gate voltage at which the Fermi energy is located at the Dirac point, $V_D$, and we typically find that $|V_D| <$ 20 V. We unambiguously determine the single layer character of a device by quantum Hall effect (QHE) measurements. Because the superconducting proximity effect requires two closely spaced electrodes, we can only perform magnetoconductance measurements in a two terminal configuration. In general, the conductance, $G$, measured in this way is a mixture of longitudinal and Hall signals, but at high fields $G \approx |G_{Hall}|$ (this approximation is exact at the Hall plateaus[19]). Indeed, the measurement of $G$ versus $V_G$ at $B = 10$ T shows clearly identifiable Hall plateaus at half-integer multiples of $4e^2/h$ (Fig. 1d), characteristic of the QHE in single layer graphene[2, 3]. This demonstrates that, even in mesoscopic samples, the quantum Hall effect can be used to identify single layer devices.

Cooling down the devices below the critical temperature of the electrodes ($T_c \sim 1.3$ K) leads to proximity-induced superconductivity in the graphene layer. A direct proof of induced superconductivity is the observation of a Josephson supercurrent[20]. Figure 2a shows the current-voltage ($I$-$V$) characteristics of a single layer device. The current flows without resistance (no voltage drop at finite current) below the critical current, $I_c$ (what we actually measure is the switching current; the intrinsic $I_c$ may be higher[20, 21]). In our devices, $I_c$ ranged from ~10 nA to more than 800 nA (at high $V_G$). Remarkably, we have measured proximity-induced supercurrents in all the devices that we tested (17 flakes in total, with several devices on some flakes), including four flakes that were unambiguously identified as single layer graphene via QHE (the rest being probably two to four layers thick). This clear observation demonstrates the robustness of the Josephson effect in graphene junctions. (The data shown are representative of the general behaviour observed; All the measurements shown have been taken on the same device, except Fig. 2b and 4, which correspond to a different single layer device, shown in Fig. 1a).

To further investigate the superconducting properties of our devices we measured the dependence of $I_c$ on magnetic field (Fig. 2b). The critical current exhibits an oscillatory

Fraunhofer-like pattern, with at least 6 visible side lobes, which is indicative of a uniform supercurrent density distribution[22]. The periodicity of the oscillations is theoretically expected to be approximately equal to a flux quantum $\Phi_o$ divided by the junction area. The area that corresponds to the 2.5±0.5 mT period is 0.8±0.2 μm$^2$, which is in good agreement with the measured device area (0.7±0.2 μm$^2$) determined from the AFM image. Applying an rf-field to the sample results in the observation of quantized voltage steps, known as Shapiro steps, in the $I$-$V$ characteristics[20]. The voltage steps, of amplitude $\hbar\omega/2e$ ($\omega$ is the microwave frequency), are a manifestation of the ac Josephson effect, and are evident in Fig. 2d. The induced superconductivity manifests itself also at finite bias in the form of subgap structure in the differential resistance due to multiple Andreev reflections[23], as shown in Fig. 2c. This subgap structure consists of a series of minima at source-drain voltages $V=2\Delta/en$ ($n=1,2…$), which enables us to determine the superconducting gap. We find $\Delta$ = 125 μeV as expected for our Ti/Al bilayers[18]. For $V > 2\Delta/e$, the differential resistance returns to the normal state value.

Having established the existence of the Josephson effect in graphene, we analyse the gate voltage dependence of the critical current. Figure 2a shows several $I$-$V$ traces taken at different $V_G$, where it can already be seen that varying the gate voltage has a strong effect on the maximum supercurrent flowing through the device. This behaviour can be more readily seen in Fig. 3a, where the differential resistance is plotted as a function of current bias and gate voltage. By changing $V_G$ we can shift continuously the Fermi energy from the valence band ($V_G < V_D$) to the conduction band ($V_G > V_D$): irrespective of the sign of $V_G$, we find a finite supercurrent. This demonstrates that the devices operate as bipolar supercurrent transistors: the supercurrent is carried by hole Cooper pairs when the Fermi level is in the valence band and by electron Cooper pairs when it is in the conduction band. Note that in going from valence to conduction band, we sweep the position of the Fermi level through the Dirac point. Strikingly, even then the supercurrent remains finite, despite the fact that for perfect graphene theory predicts a vanishing density of states at $V_G=V_D$[24]. This behaviour has been observed in all samples and demonstrates that electronic transport in graphene is phase coherent irrespective of the gate voltage, including when the Fermi level is located at the Dirac point.

In conventional Josephson junctions, the critical current correlates with the normal state conductance, $G_n$[25]. In graphene this correlation can be observed directly, as shown in Fig. 3a, because both $I_c$ and $G_n$ depend on $V_G$. To analyse this correlation, we plot the product of the measured critical current and the normal state resistance ($R_n = 1/G_n$), or $I_cR_n$ product (see Fig. 3b). We find that at high gate voltage $I_cR_n \approx \Delta/e$, and that $I_cR_n$ is suppressed by a factor 2-3 around the Dirac point. The observed $I_cR_n$-product is thus close to the theoretical value of ~2.5$\Delta/e$ for a model system of graphene in the ballistic regime [15]. The remaining discrepancy could be easily accounted for by the difference between the measured critical current (or switching current) and the intrinsic critical current. [20, 21] This difference is in general hard to quantify and depends on the electromagnetic environment of the sample, but it is expected to be more pronounced when $I_c$ itself is smaller (i.e. at low $V_G$), and could thus well explain the observed $V_G$ dependence of $I_cR_n$.

An interesting aspect of supercurrent in graphene is the special role of time-reversal symmetry in the Josephson effect compared to that in phase coherent transport in the normal state. At low energy, the band structure of graphene consists of two identical, independent valleys centred on the so-called $K_1$ and $K_2$ points[24], which transform into each-other upon time reversal. Since Cooper pairs are made out of time reversed electron states, the two electrons in Cooper pairs that are injected from the superconducting electrodes into graphene go to opposite K-points[15] (see Fig. 1b). In this way, the presence of a superconducting electrode provides an intrinsic mechanism that couples phase coherently electronic states belonging to opposite valleys. The dynamics of electrons is then described by the full (two-valley) Hamiltonian of graphene that is time-reversal symmetric.

The situation is very different for normal-state transport where the dynamics of the electrons is determined by a "one-valley" Hamiltonian. This has not only important consequences for the QHE,[4,5] but also for quantum interference. When the two valleys are fully decoupled, interference originating from coherent propagation of electrons along time reversed trajectories cannot occur, as these trajectories involve quantum states from opposite K-points. As a consequence, quantum corrections to the conductivity, like weak localization, would be absent (note that for the single-valley Hamiltonian there exists an effective TRS in the long wavelength limit, but in real samples this symmetry is easily broken[26-28]. See supplementary information). The contribution of time-reversed trajectories to quantum interference can be restored only up to the extent that there exists a coupling mechanism between the two valleys such as short-range scattering at impurities or at the graphene edges[26-28]. This makes the occurrence of weak localization dependent on extrinsic defects, which explains why both in our own samples (see Fig. 4b) and other recent measurements[17,29] weak localization is found to exhibit a surprising sample dependent behaviour and is often suppressed (while, at the same time, we observe a supercurrent in all samples investigated). This interpretation, which illustrates the unique electronic properties of graphene, is consistent with the observation of aperiodic conductance fluctuations[30] of amplitude $e^2/h$ (see Fig. 4a), whose occurrence requires phase coherence but not time reversal symmetry.

Supplementary Information is linked to the online version of the paper at www.nature.com/nature.

**Author Information:** Reprints and permissions information is available at npg.nature.com/reprintsandpermissions. The authors declare no competing financial interests. Correspondence and requests for materials should be addressed to H.B. (hubert@kavli.nano.tudelft.nl) or P.J. (P.D.Jarillo-Herrero@tudelft.nl).

We gratefully acknowledge A. Geim, D. Jiang , and K. Novoselov for help with device fabrication; L. Kouwenhoven for the use of experimental equipment, support and discussions; and C. Beenakker, J. van Dam, D. Esteve, T. Klapwijk, Y. Nazarov, G. Steele, B. Trauzettel, C. Urbina, and B. van Wees for helpful discussions.

**Figure 1.** Sample characterization. **a,** Atomic force microscope image of a single layer graphene device in between two superconducting electrodes. We have fabricated devices with electrode separations in the range 100-500nm. **b,** Schematic representation of graphene in between superconducting electrodes. The two electrons in a Cooper pair entering graphene go into different K-valleys. **c,** Two terminal resistance versus gate voltage, $V_G$, at $T = 30$ mK and a small magnetic field, $B = 35$ mT, to drive the electrodes into normal state. The aperiodic conductance fluctuations are due to random quantum interference of electron waves (see also Fig. 4). **d,** Two terminal conductance, $G$, versus $V_G$ at high magnetic field, $B = 10$ T, and $T = 100$mK, showing a series of steps at half-integer values of $4e^2/h$, characteristic of the anomalous quantum Hall effect in single layer graphene.

**Figure 2.** Josephson effect in graphene. **a,** Voltage, $V$, versus current bias, $I$, characteristics at various $V_G$, showing a modulation of the critical current. Inset: current bias sweeps from negative to positive (red) and viceversa (blue), showing

that the asymmetry in the main panel is due to hysteretic behaviour (the retrapping current is smaller than the switching current, as is typical for an underdamped junction). **b,** Color-scale representation of *dV/dI*(*I*, *B*) at *T*=30mK (yellow-orange is zero, i.e. supercurrent region, and red corresponds to finite *dV/dI*). The critical current exhibits a series of oscillations described by a Fraunhofer-like pattern. **c,** Differential resistance, *dV/dI*, versus *V*, showing multiple Andreev reflection dips below the superconducting energy gap. The dips in *dV/dI* occur at values of $V = 2\Delta/en$, where *n* is an integer number. **d,** ac-Josephson effect. The Shapiro steps in the *I-V* characteristics appear when the sample is irradiated with microwaves. In the example, we applied 4.5GHz microwaves, resulting in 9.3µV voltage steps. Inset: Colour-scale plot showing the characteristic microwave amplitude ($P^{1/2}$) dependence of the ac Josephson effect (orange/red corresponds to zero/finite *dV/dI*).

**Figure 3.** Bipolar supercurrent transistor behaviour and finite supercurrent at the Dirac point. **a,** Colour-scale plot of *dV/dI* (*V$_G$*,*I*). Yellow means zero, i.e., supercurrent region, and finite *dV/dI* increases via orange to dark red. The current is swept from negative to positive values, and is asymmetric due to the hysteresis associated with an underdamped junction (see also inset to Fig. 2a). The top axis shows the electron density, *n*, as obtained from geometrical considerations[1]. For large negative (positive) *V$_G$* the supercurrent is carried by hole (electron) Cooper pairs. The supercurrent at the Dirac point is finite. Note that the critical current is not symmetric with respect to *V$_D$*. The origin of this asymmetry is not known, but a simmilar asymmetry is seen in the normal state conductance (blue curve). **b,** Product of the critical current times the normal state resistance versus *V$_G$*. The normal state resistance is measured at zero source-drain bias, at *T* = 30mK and with a small magnetic field to drive the electrodes in the normal state. The $I_c R_n$ product exhibits a dip around the Dirac point (see main text).

**Figure 4.** Magnetoconductance measurements. **a,** Conductance versus magnetic field at $T \sim 1.5K$ for a single layer device at $V_G = V_D$. The red and black traces are two subsequent measurements, showing reproducible conductance fluctuations, whose root mean square amplitude is $e^2/h$. **b,** Low field magnetoconductance measured at $T \sim 1.5K$ (electrodes in the normal state) for two different single layer devices (the red trace corresponds to the device measured in **a**), showing that the amplitude of the weak localization effect is sample dependent and supressed as compared to the expected value $\sim e^2/h$. Each curve results from an ensemble average of 66 individual magnetoresistance measurements taken at different gate voltages near the Dirac point.

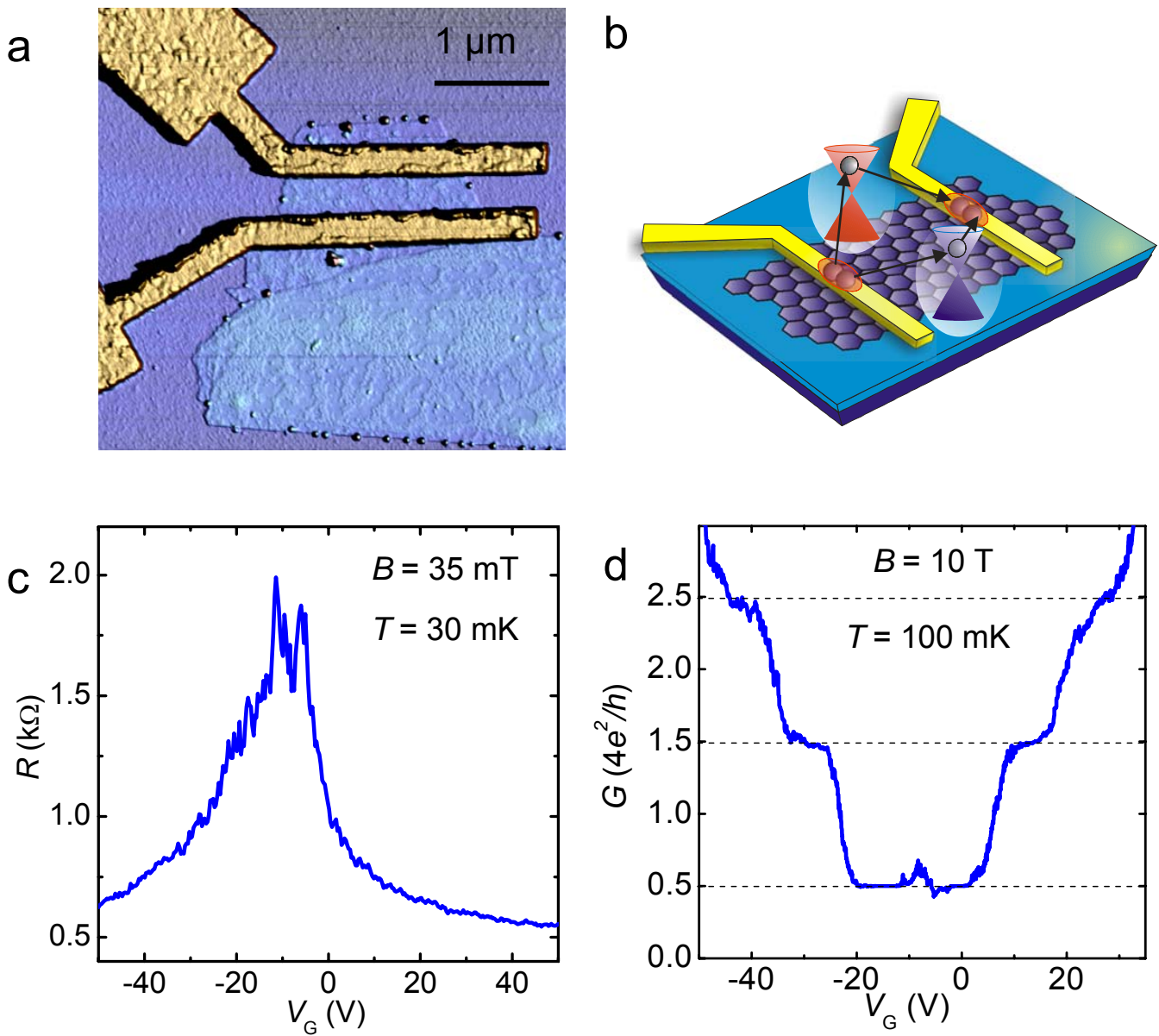

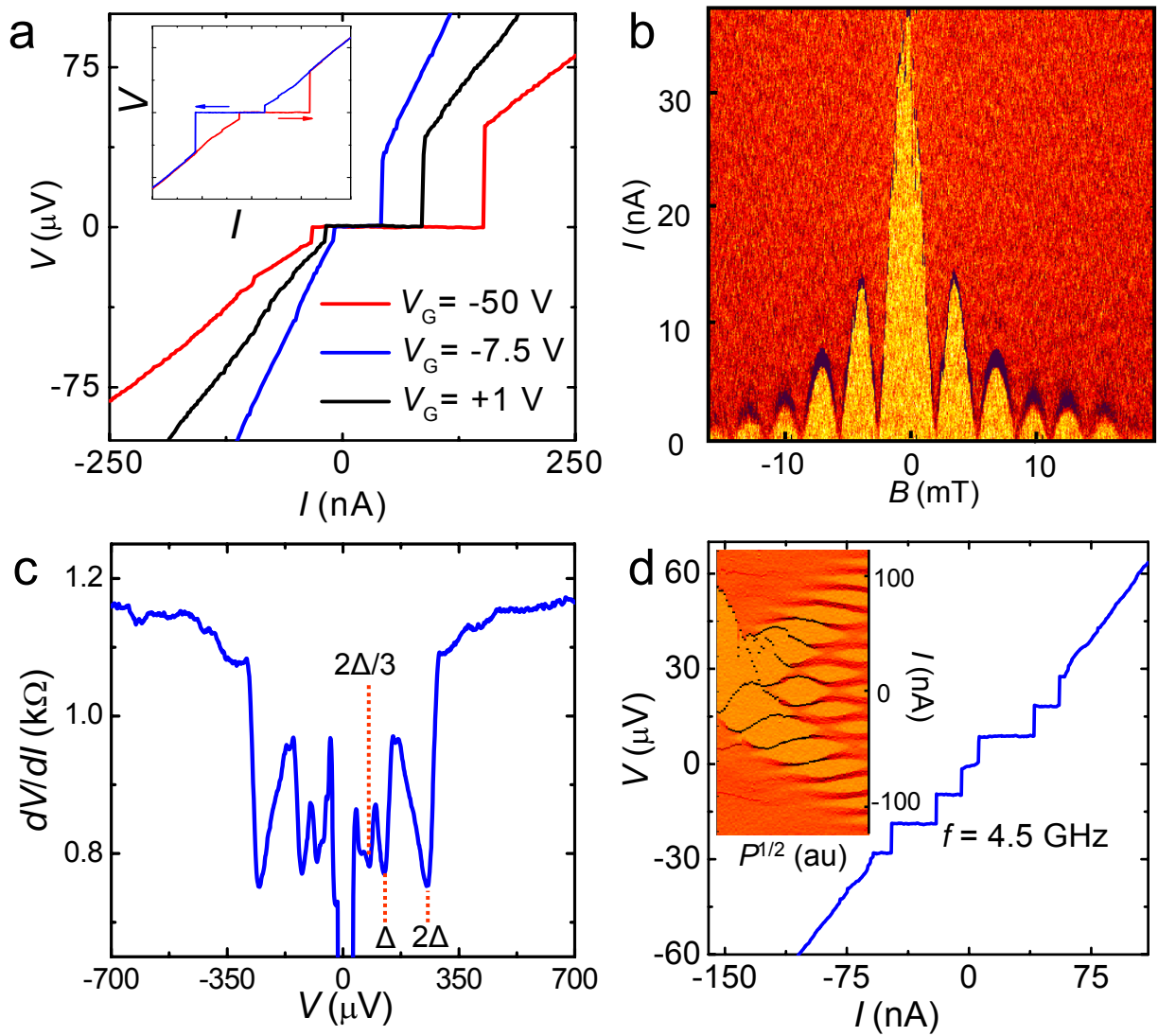

Figure 2 Heersche *et al.*

Figure 3 Heersche *et al.*

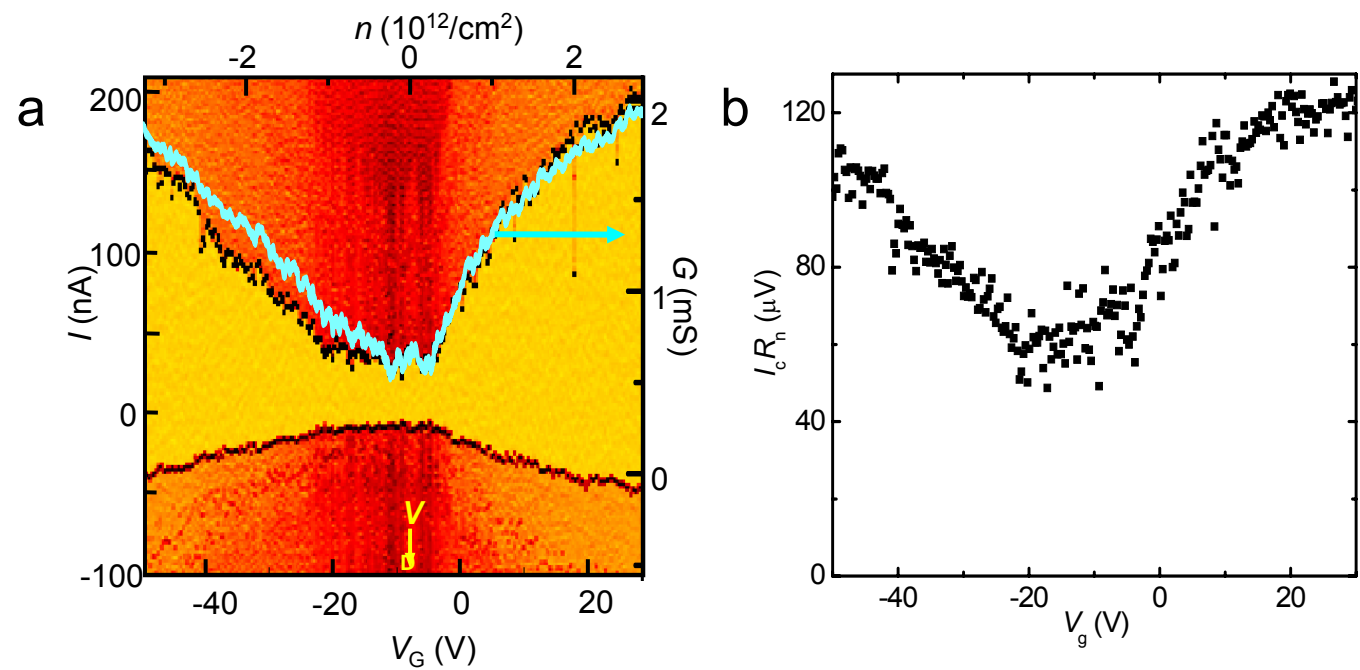

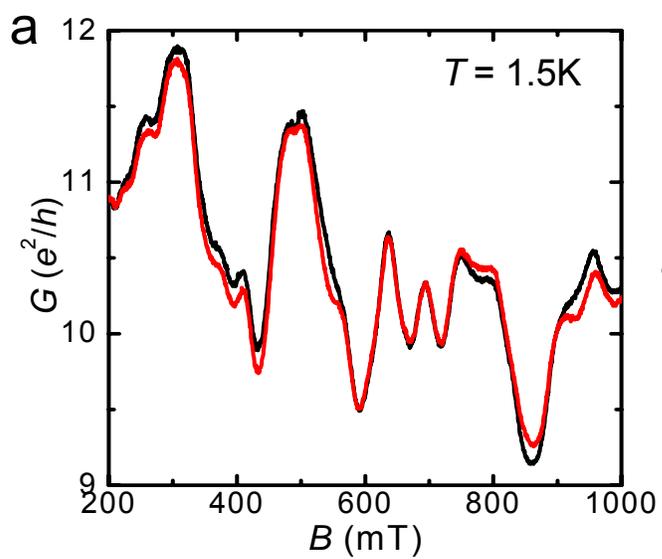
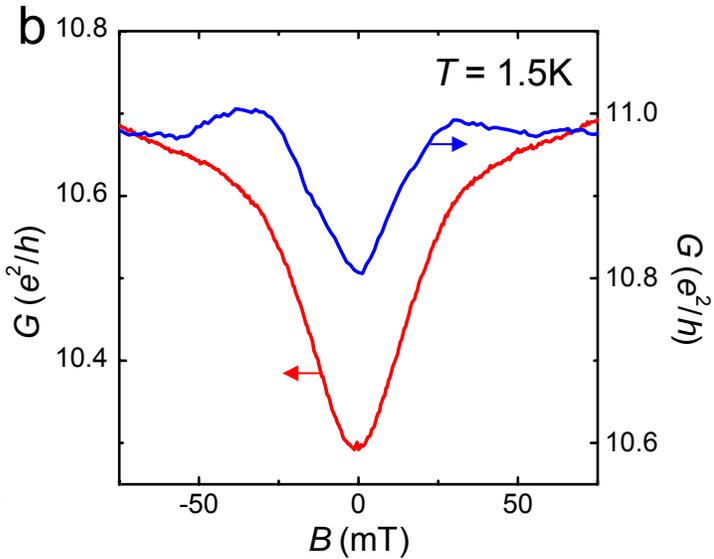

# Supplementary information

**Bipolar supercurrent in graphene**

Hubert B. Heersche, Pablo Jarillo-Herrero, Jeroen B. Oostinga, Lieven M. K. Vandersypen and Alberto F. Morpurgo

**1. Effective time reversal symmetry breaking in graphene**

In the long wavelength limit, the effective Hamiltonian for the valley around one K-point is a 2x2 matrix acting on the components of the wavefunction on the two independent sublattices of graphene (the so-called A and B atoms). It reads

$$H_K = \hbar v_F \begin{pmatrix} 0 & k_x - ik_y \\ k_x + ik_y & 0 \end{pmatrix}$$

with $\vec{k} = (k_x, k_y)$ the momentum of the electron measured relative to the K-point. This 2x2 matrix is left invariant by the anti-unitary transformation $U = i\sigma_y \hat{K}$, i.e. $U^\dagger H_K U = H_K$, where $\sigma_y$ is the usual Pauli matrix and $\hat{K}$ denotes complex conjugation. *U* is the usual operator associated to the time reversal symmetry transformation for spin ½ particles. In this sense, the operator *U* defines an "effective" time reversal symmetry transformation that is applicable in a single valley only and reverses the sign of the momentum $\vec{k}$ relative to the K-point.

In the presence of certain types of smooth scattering potentials (see Ref. [1]), there exist "effectively" time-reversed trajectories along which the electrons can propagate while remaining within the same valley. On this basis, Suzuura and Ando[1] predicted that these trajectories cause a quantum correction to the conductivity in the form of weak antilocalization.

The operator *U* however does not represent the true time reversal symmetry transformation for electrons in graphene, as true TRS connects states belonging to the two opposite K-points in graphene and cannot be described within the single K-point approximation. For this reason the effective TRS symmetry represented by *U* is not robust against physical perturbations that normally preserve TRS (e.g. static potentials). For instance, effective TRS is broken by (i) curvature in graphene[2,3] (that can be described within a single K-point approximation as a vector potential), (ii) scattering at edges[4], and (iii) trigonal warping[5] (i.e., by including quadratic terms in the long wavelength expansion from which the one K-point Hamiltonian is obtained). As a consequence, in real experimental samples the observation of effects associated to the symmetry *U* is expected to be very difficult (at least in samples of current quality).